# Secure Tracking in Sensor Networks using Adaptive Extended Kalman Filter

Ali P. Fard
*Department of Electrical Engineering, Sharif University of Technology, Tehran, Iran.*

Mahdy Nabaee
*Department of Electrical and Computer Engineering, McGill University, Montreal, Qc, Canada*

**Abstract**

*Location information of sensor nodes has become an essential part of many applications in Wireless Sensor Networks (WSN). The importance of location estimation and object tracking has made them the target of many security attacks. Various methods have tried to provide location information with high accuracy, while lots of them have neglected the fact that WSNs may be deployed in hostile environments. In this paper, we address the problem of securely tracking a Mobile Node (MN) which has been noticed very little previously. A novel secure tracking algorithm is proposed based on Extended Kalman Filter (EKF) that is capable of tracking a Mobile Node (MN) with high resolution in the presence of compromised or colluding malicious beacon nodes. It filters out and identifies the malicious beacon data in the process of tracking. The proposed method considerably outperforms the previously proposed secure algorithms in terms of either detection rate or MSE. The experimental data based on different settings for the network has shown promising results.*

## 1. Introduction

The development of small, low-power, low-cost and multifunctional sensor nodes have offered viable solutions for wide range of applications. Various applications have been proposed for such networks and a number of them have been already deployed. Many of these applications require the knowledge of the location of the nodes. Environmental monitoring, emergency rescue, target detection and military surveillance and tracking need to know location information as a priori. Location information is also a necessity in different operations of networks such as routing and security protocols [4]. To fulfill this emerging demand numerous algorithms have been proposed to provide accurate location information.

Despite the extensive studies conducted on localization of sensors in trusted environments, few have considered this process in hostile environments. Since the sensors may be deployed in unsupervised manner they are susceptible to many security threats trying to corrupt the proper function of localization process. Many of these threats are not the conventional security threats and they target the reliability of location information instead [5]. Since localization techniques mainly use different features of radio signals of devices as their input data, e.g. Time of Arrival (TOA) and Received Signal Strength (RSS), an adversary can easily influence the measurements by jamming and replying or modifying the signal strengths. Hence, the performance of location estimation techniques is severely ruined when used in hostile environments.

Secure localization problem has been addressed in different variants. Some has enhanced the robustness of localization process in hostile environments and some has gone for verification of the estimated location before they are further incorporated into the process of localizing other nodes.

Several techniques have been proposed to securely localize sensor nodes in hostile environments [6], [7], [8], [9] but to our best knowledge secure tracking of mobile objects in WSNs has received very little attention.

Conventional tracking algorithms incorporate Bayesian filtering, particle filters or EKF filters. For higher accuracy of the estimated path, different motion models may be defined to account for more intricate movements [6]. Some solutions to tracking problem based on Kalman Filtering have been suggested in [1], [10] and [11], but none of them has considered the security of the procedure. When false locations are injected into network they fail to provide the correct solution and in most cases diverge.

For cost effectiveness, sensor nodes in WSN estimate their location based on the information they obtain from nodes with fixed known locations, called Anchor nodes, which are usually supposed to be tamper proof. Anchors are, however, highly vulnerable to be subverted during operation phase, the consequence of which would be false data being injected into network and hence the process of tracking will not come up with the correct trajectory for the MN. In this paper, we formulate a secure accurate tracking algorithm which is able to track the object under study in the presence of high measurement noise and compromised anchor nodes. It is able to identify and filter out the misbehaving anchors. The remainder of this paper is as follows. Section 2 reviews the state of the art in tracking and the attempts made to provide robuster algorithms. In section 3 we formulate the state and measurement model. The process of removing the malicious anchors is described in section 4 and simulation results are explained in section 5.

## 3. Related work

Reliable estimation location estimation of nodes suffer from both inherent errors imposed by transmission medium and attacks conducted to deliberately change the physical medium and hence the input data to localization algorithms. As for the limited resources available in WSNs, traditional security techniques based on cryptography cannot be used as the main defense. In fact the advantage of non-cryptographic algorithms is better understood in the presence of compromised nodes where the attacker can authenticate itself and the integrity checks fail to identify erroneous packets.

In [6] a rang-independent localization algorithm with a special framework has been introduced which provides protection against most of the attacks on distance estimation. [4] has proposed methods for identification and robust computation of location by comparing the MMSE of sensors location information with some consistent nodes and further utilizing a voting-based algorithm, respectively. In [5] least square method is used to localize nodes and least median square method when the network is determined to be under attack. Although the mentioned methods and several more has been proposed for robust positioning of nodes, little work has been done for secure tracking of network nodes.

The secure tracking based on Bayesian Filtering in [2] proposes to activate 3 adjacent nodes for tracking the MN at each time step and further use a relaxation labeling algorithm to identify and remove the false data by malicious nodes. The malicious nodes are identified if the Euclidean of their calculated belief for the state of the system is drastically different for those of benign ones. The MN is supposed to have a linear motions model and the attack is simulated by assuming the reports of the malicious nodes being located on a predetermined artificial path. To assess the algorithm the proposed relaxation labeling method is compared with simply averaging the results of particle filtering for each of the 3 nodes at each time step. 15 sensor nodes are deployed within [0,30] and MSE of the results compared to the true target path with one or two attackers are 12.107, 13.382. Another secure tracking is also explained in [3], which supposes the MN to be pseudo-static and tracks its path through iterative application of a localization algorithm. This, however, imposes more complexity than conventional tracking techniques. It calculates the position of the target as the intersection of the circles centered at the anchors with its radius equal to the estimated distance by that sensor. The attacks are simulated only by distance enlargement and an anchor is determined to be malicious if its reported position of the target is different from the estimated one. It has also determined that the number of trustful anchors for correct localization should be at least equal to the number of malicious anchors plus four.

## 3. Target Tracking

### 3.1. System model assumptions

1) It is assumed that the network consists of at least one base station and a set of uniformly deployed anchors with fixed known locations.
2) Malicious anchor nodes are capable of establishing their authenticity with the rest of the sensors and have access to the shared symmetric keys.
3) Received Signal Strength (RSS) values of sensor nodes are used as the input data for the process of tracking.
4) Sensor nodes other than anchors will estimate their location based on the measurements provided by anchors.
5) The malicious anchor nodes, which are able to collude, will report enlarged or reduced locations to the base station.

### 3.2. Tracking technique

The measurements obtained from sensor nodes are used in Extended Kalman Filter to arrive at the state of the system. The process is governed by the following nonlinear difference equation:

$x_k = f(x_{k-1}, w_{k-1}), \quad w_k \sim N(0, Q_k)$ (Eq. 1)

Where $x_k$ represents the state vector in the $k^{th}$ moment and $w_k$ is the corresponding process noise. The measurement equation is:

$z_k = h(x_k, v_k), \quad v_k \sim N(0, R_k)$ (Eq. 2)

In which $z_k$ and $v_k$ are the measurement vector and its noise in the $k^{th}$ moment, respectively. $R_k$ represents the measurement noise covariance which reflects the quality of the received measurements.

The empirically verified equation used to model the measured power (RSS) and the corresponding distance ($d$) between the transmitter and the receiver is

$P_r[dB] = P_0[dB] - \alpha \cdot \log_{10}\left(\frac{d}{d_0}\right)^2 + n$ (Eq. 3)

Where α is the path loss parameter, $P_r$ and $P_0$ are the received and emitted powers. As it is mentioned in [1], ($n$) is the log-normal shadow fading which is modeled as a normal distribution with zero mean.

Taking the reference distance $d_0 = 1m$ the received signal strength can be modeled as:

$P_{r,i} = P_0 - \alpha \log_{10}[(x^k - x_i)^2 + (y^k - y_i)^2] + n$ (Eq. 4)

In this equation, $(x^k, y^k)$ is the coordinate of MN in the $k^{th}$ moment. The coordinate of $i^{th}$ anchor node is $(x_i, y_i)$ and the corresponding measured RSS is represented by $P_{r,i}$.

The adopted measurements model is the one used in [1]. Thus, the measurements, $z_i^k$, are as follows:

$z_i^k = 10^{-(P_i^k - P_0)/\alpha}$ (Eq. 5)

Reformulating this equation we obtain the measurement model used

$z_i^k = [(x^k - x_i)^2 + (y^k - y_i)^2] \times 10^{-n/\alpha}$ (Eq. 6)

The EKF time-update equations will be as follows:

Table 1. EKF time-update equations

$\hat{x}_k^- = A \cdot x_{k-1}$
$P_k^- = A \cdot P_{k-1} \cdot A^T + Q$

And the measurement-update equations are shown in Table 2.

Table 2. EKF measurement-update equations

$K_k = P_k^- \cdot H_k^T \cdot (H_k \cdot P_k^- \cdot H_k^T + V_k \cdot R_k \cdot V_k^T)^{-1}$
$\hat{x}^k = \hat{x}_k^- + K_k \cdot (z_k - (\hat{d}_k^-)^2)$
$P_k = (I_4 - K_k \cdot H_k) \cdot P_k^-$

$A = \begin{bmatrix} 1 & 1 & 0 & 0 \\ 0 & 1 & 0 & 0 \\ 0 & 0 & 1 & 1 \\ 0 & 0 & 0 & 1 \end{bmatrix}$ (Eq. 7)

In the preceding equations, $x^k = [x^k \; \dot{x}^k \; y^k \; \dot{y}^k]^T$ represents the state vector of the system which we assumed to be the coordinates of the source node along with the corresponding velocities in two dimensional axes and thus $A$ would be as shown in Eq. 7.

As the measurement equation is nonlinear it should be linearized around predicted target state. Here H is the Jacobian matrix of partial derivatives of $h(.)$ with respect to $x$ that in fact indicates what the measurements can say about the state vector.

$H_{[i,j]} = \frac{\partial h_{[i]}}{\partial x_{[j]}}(\hat{x}_k, 0)$ (Eq. 8)

V is the Jacobian matrix of partial derivatives of $h(.)$ with respect to:

$V_{[i,j]} = \frac{\partial h_{[i]}}{\partial v_{[j]}}(\hat{x}_k, 0)$ (Eq. 9)

Whose elements will be [1]:

$\begin{vmatrix} h_{i,1}^k = 2(\hat{x}^{k-1} - x_i) \\ h_{i,3}^k = 2(\hat{y}^{k-1} - y_i) \\ h_{i,2}^k = h_{i,4}^k = 0 \end{vmatrix}$ (Eq. 10)

and,

$\begin{vmatrix} v_{i,i}^k = -\frac{\ln 10}{\alpha} \cdot (d_i^k)^2 \\ v_{i,j}^k = 0 \end{vmatrix}$ (Eq. 11)

## 4. Detection of Malicious Sensors

As the measurements of anchors are used to estimate the position of sensor nodes, the presence of false data will greatly affect the predicted trajectory and the process of filtering will diverge most of the time. So, the proper operation of the previously proposed algorithms of tracking is very dependent on the detection and discarding the false data.

False measurements will change the Kalman gain introduced in table 2 and as a result the posterior values of the state of the system will not be valid. Thus, it is tried to identify and remove the sensors with malicious activities during the process of Kalman Filter. The introduction of false data will also introduce a lot of change in the subsequent values of the estimated state variables. This information is very useful in detection of the misbehaving sensors and is utilized as described below.

Significant difference was observed between the values of the distance among the target and the sensors based on the estimated location and RSS. This fact was utilized to define a condition based on which malicious anchors were identified. A priori distance between the MN and the $i^{th}$ sensor, $Dist_i^-$, is calculated from the corresponding RSS. Its a posteriori estimate, $Dist_i^+$, is obtained as the result of the process of estimation. The following equations formulate their relation with the measured RSS and the estimated location.

$$Dist_i^- = 10^{-(P_{r,i}-P_0)/\alpha} \qquad \text{(Eq. 12)}$$

$$Dist_i^+ = \sqrt{(\hat{x}^k - x_i)^2 + (\hat{y}^k - y_i)^2} \qquad \text{(Eq. 13)}$$

And the difference of these two distances, $\Delta_i^k$, was observed to be a good measure to detect the malicious nodes. This is because the false reported location of the sensor or RSS value by the malicious node will cause a significant difference between the values of $Dist_i^-$ and $Dist_i^+$. Thus we may define $\Delta_i^k$ as follows:

$$\Delta_i^k = |Dist_i^- - Dist_i^+| \qquad \text{(Eq. 14)}$$

An example of the resulting $\Delta_i^k$s for different nodes of a network in which one of the nodes is transmitting its location wrongfully is depicted in Fig. 1. In this figure, the dotted line represents the average value of $\Delta_i^k$s in each time, $\Delta_{av}^k$, defined by Eq. 15. This value is used for detection of malicious nodes and refining the reported measurements.

$$\Delta_{av}^k = \frac{1}{N_s} \sum_{i=0}^{N_s - 1} \Delta_i^k \qquad \text{(Eq. 15)}$$

Where $N_s$ is the number of sensors in the network.

It was observed that higher noise will result in rejection of some honest nodes too. To prevent false rejection of nodes the distribution of the mean of this error when there was no malicious node was studied and a threshold on error of measurement for honest nodes was derived.

To obtain the suitable condition many probability functions were studied. The following was proved to be a good condition, based on which the misbehaving nodes could be rejected while the honest nodes are not reported as a malicious node in high noise deployments. The detection process is to test the following condition:

$$\Delta_i^k > \Delta_{av}^k + \gamma \times (\Delta_{max}^k - \Delta_{min}^k) \qquad \text{(Eq. 16)}$$

Where $\Delta_{max}^k = \max_i \Delta_i^k$ and $\Delta_{min}^k = \min_i \Delta_i^k$. Clearly, if Eq. 16 holds then the corresponding node is detected as a malicious node. Afterwards, the measured RSS and reported sensor location related to that node is removed from the set of network nodes. A statistical parameter called $\gamma$ is put in the preceding condition. This parameter is derived by statistical study on different deployments of sensors with different measurement noises. This condition enabled us to detect one malicious node among 6 anchors in 78% of the times. It should be noted that as the initial start of detection process is very dependable on the initial conditions for prediction and update phases. If the initial state estimate was not accurate, it will take some iteration for the filtering process to be able to track the MN's trajectory. This will introduce enough errors that the anchors may be detected misbehaving. Thus a window is defined after which the process of detection will start. With this simple classifier, detection of malicious nodes is done after each iteration of the EKF tracking and the tracking process will be performed again after each of the malicious nodes were detected and removed from the set of network nodes.

Considering the linear relationship between the target state and the measurements as defined in Eq. 17, the innovation vector ($v_k$) is defined as the difference between the measurements and the predicted target position ($z_k - H \cdot x_k$).

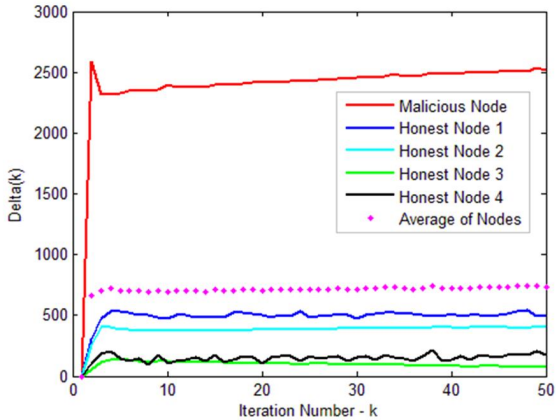

**Figure 1** $\Delta_i^k$ for different nodes and their average

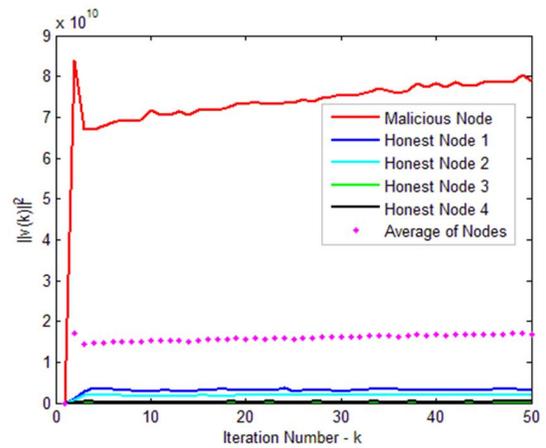

**Figure 2** $\|v_k\|^2$ for different nodes and their average.

$$z_k = H \cdot x_k + v_k \qquad \text{(Eq. 17)}$$

Here $\Delta_i^k$ is in fact the innovation vector for the $i^{th}$ sensor. The innovation vector which is a natural measure of the discrepancy between the actual measurement and the predicted measurement is assumed to be normally distributed ($N(0, C_k)$).

Further investigation of the problem revealed that the deviation between the predicted and actual measurement is reflected much better with Eq. 18:
$$\|v_k\|^2 = v_k^T C_k^{-1} v_k \quad \text{(Eq. 18)}$$
This measure, also called Mahalanobis norm, was proved a more efficient statistical classifier since it takes the correlation of the measured data into account. The real value of $C_k$, the covariance of the innovation vector, is estimated after a window of 10 received measurements.

It is shown in Fig. 2 with the same condition as Fig.1. As it's seen the malicious node can be discriminated better.

A threshold based on the new criterion was utilized as before to prevent false rejection of the honest nodes. The Mahalanobis distance is, however, scale invariant as opposed to Euclidean distance and because of this the threshold remained constant under different conditions. The sensor nodes were detected as malicious if the Mahalanobis distance of its reported measurements where higher than the average value. As before the process of detection will be done in each iteration of the EKF tracking and once the malicious nodes are detected they will be removed and the tracking procedure will be performed again.

The experimental results demonstrated that with the same deployment the malicious node could be identified above 85% of times even with higher noise powers.

## 5. Simulation Results

Sensor nodes were uniformly deployed in a $100m \times 100m$ rectangular region. The target is assumed to follow a straight path and the RSS values are modeled as described in (Eq. 4) where the noise power is assumed to be 0.5dB. The attack was simulated by randomly selecting a number of anchors. Malicious anchors change the reported locations with a normal noise with zero mean and $\sigma_n = 20$. The network has one malicious anchor and 5 honest anchors. The effect of varying the number of malicious nodes, anchor nodes and noise power is studied.

Here the 'true detection' refers to the rate of malicious anchors being detected and 'false detection' is occurred when an honest anchor is erroneously detected to be misbehaving.

Fig. 3 shows the number of malicious nodes being detected with different number of anchors available for the process of tracking. As shown in Fig. 3 remarkably high percentage of malicious nodes are detected.

Fig. 4 shows the obtained MSE as the density of the honest anchors is changed. One node is randomly set to provide false data. As it is seen that the malicious node is detected with high probability and even with the presence of the malicious anchor this methods obtains very low MSE.

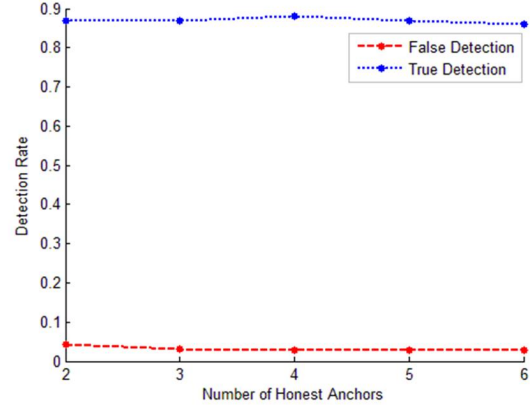

**Figure 3** Rate of detection of malicious nodes. There is one malicious anchor and 5 honest anchors.

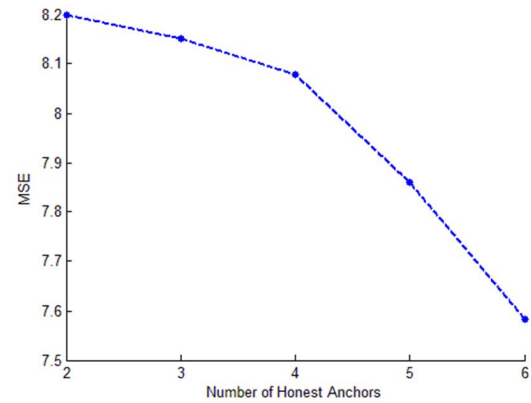

**Figure 4** MSE of tracking for different No. of sensors. There is one malicious anchor and 5 honest anchors.

The effect of more malicious anchors on the rate of detection is shown in Fig. 5.

Here a total of 6 anchors are supposed. It was observed that as the number of malicious anchors where increased the rate of true detection where decreased, though the MSE of algorithm has maintained a low amount. As seen in Fig. 6 even with half of the anchors being malicious this method has good performance and the algorithm was able to detect many of the misbehaving nodes. Hence, the honest nodes should not necessarily outnumber the malicious nodes and the bounds that may be imposed on the number of malicious anchors for proper operation is higher than previously proposed methods.

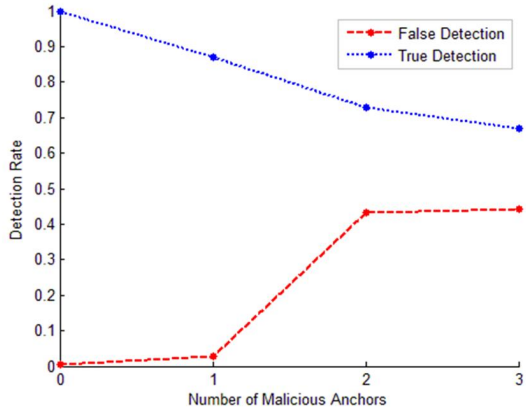

**Figure 5** Rate of detection of malicious nodes. There are a total of 6 anchors present.

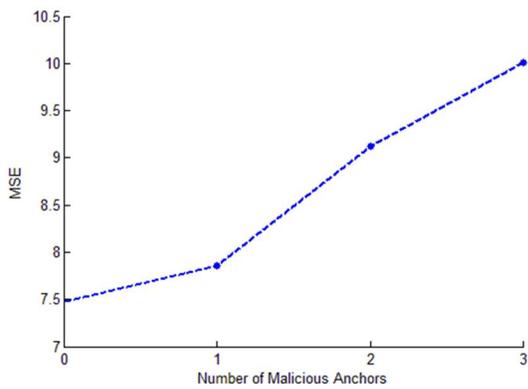

**Figure 6** MSE of Tracking for different number of malicious nodes. There are a total of 6 anchor nodes.

It is worth mentioning that although false detection of sensor nodes generally seems to degrade the performance, it will improve the performance when higher number of anchors is deployed. This is clearly due to the fact that false detection of an anchor is because of the high noise of its measurement, the removal of which will improve the performance.

The effect of the location error introduced by malicious anchors is studied in Fig. 8. As it's seen the MSE has maintained almost a constant value, however, the detection rate may be worse when the introduced error is not significant. This is obviously due to the fact that with smaller errors the MSE will not deviate much from its expected value.

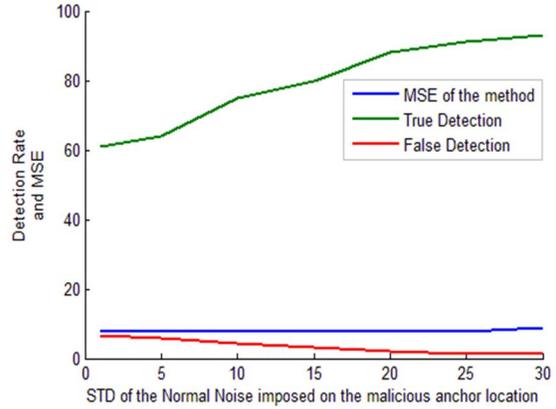

**Figure 8** Rate of Detection and MSE of tracking vs. the STD of the noise introduced by malicious anchors. There is one malicious anchor and 5 honest nodes.

The effect of higher noise on the performance of the system is shown in Fig. 7, 8. Obviously higher noise power will degrade system performance but this had very little effect on performance in terms of detection. Here, as expected, noisier deployment along with slightly degraded detection rate has increased the MSE.

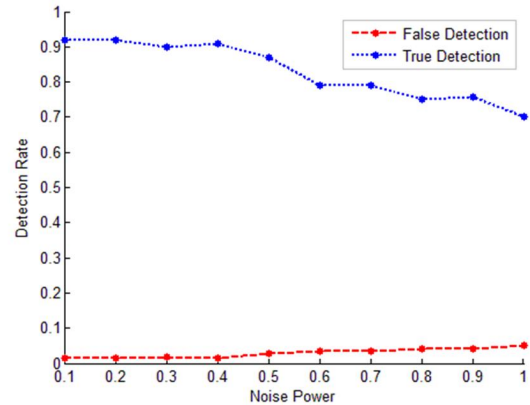

**Figure 8** Rate of Detection for different Noise Powers. There is one malicious anchor and 5 honest anchors.

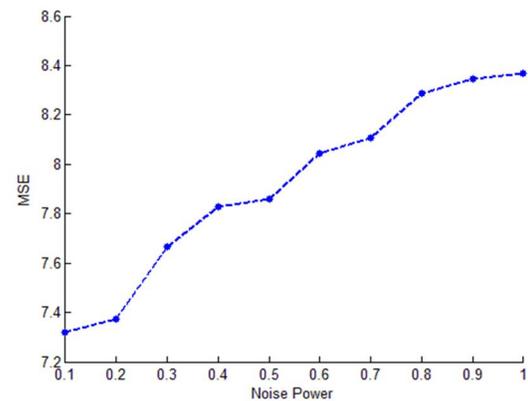

**Figure 9** MSE of Tracking for different Noise Powers. There is one malicious anchor and 5 honest anchors.

## 6. Conclusions

Many algorithms have been proposed to estimate the trajectory of a MN in sensor networks. Nevertheless few of which have considered the probability of attacks on the tracking procedure. The presence of malicious sensors will greatly degrade the performance of such algorithms and the conventional security mechanisms fail in the presence of compromised sensors. In this paper we have presented an efficient tracking algorithm applicable to hostile environments. The false measurements that malicious nodes inject into network are removed and the misbehaving anchors are identified. This algorithm was evaluated and was proved to provide noticeably good performance in terms of rate of detection of the malicious sensor nodes, MSE and the low number of honest sensors needed for proper operation of the algorithm.

## 7. References

[1] M.Nabaee, A.Pooyafard, A. Olfat, "Enhanced Object Tracking with Received Signal Strength using Kalman Filter in Sensor Networks", The 4$^{th}$ International Conference on Telecommunication (IST2008) August 27-28, 2008-Tehran, Iran.
[2] Chang, C.-C.G. Snyder, W.E. Wang, C., "Secure Tracking in Sensor Networks", IEEE International Conference on Communications, 2007. (ICC '07), 24-28 June 2007.
[3] Misra, S.Bhardwaj, S.Guoliang Xue, "ROSETTA: Robust and Secure Mobile Target Tracking In A Wireless Ad Hoc Environment", Military Communications Conference (MILCOM), 23-25 Oct. 2006.
[4] Attack-Resistant Location Estimation in Sensor Networks
[5] Z.Li, W.Trappe, Y.Zhang and B.NATH , "Robust Statistical Methods for Securing Wireless Localization
in Sensor Networks", IEEE International Conference on Communications, ICC 2007.
[6]"Joint multiple-target tracking and classification"
[7] L. Lazos, R. Poovendran, "HiRLoc: high-resolution robust localization for wireless sensor networks", IEEE Journal on Selected Areas in Communications, Volume 24, Issue 2, Feb. 2006 Page(s): 233 – 246.
[8] S. Capkun, S. Ganeriwal, F. Anjum and M. Srivastava, "Secure RSS-based Localization in Sensor Networks", Technical Reports 529, ETH Zürich, 09 2006.
[9] S. Capkun and J.-P. Hubaux, "Secure positioning of wireless devices with application to sensor networks", IEEE Infocom, pp. 1917–1928, March 2005.
[10] Chin-Liang Wang, Yih-Shih Chiou, "An Adaptive Positioning Scheme Based on Radio Propagation Modeling for Indoor WLANs,"
[10] S. Umesh Babu, C. S. Kumar, R. V. Raja kumar, "Sensor Networks for Tracking a Moving Object using Kalman Filtering,," IEEE International Conference on Industrial Technology (ICIT), Mumbai, 15-17 Dec. 2006.